\documentclass[12pt]{article}
\usepackage{cite}
\usepackage{pifont}
\usepackage{hyperref}
\usepackage{graphicx}
\graphicspath{{./}{./figures/}}

\newcommand{\vep}{\varepsilon}
\newcommand{\ep}{\epsilon}
\newcommand{\MSbar}{\overline{\mbox{MS}}}
\newcommand{\msbar}{\overline{\mbox{\scriptsize MS}}}
\newcommand{\ARI}{\mbox{RI/MOM}}
\newcommand{\ari}{\mbox{\scriptsize RI/MOM}}
\newcommand{\RIp}{\mbox{RI}^\prime\mbox{/MOM}}
\newcommand{\rip}{\mbox{\scriptsize RI}^\prime\mbox{\scriptsize/MOM}}
\newcommand{\RISMOM}{\mbox{RI/SMOM}}
\newcommand{\RISMOMB}{\mbox{RI/SMOM}_{\gamma_{\mu}}}
\newcommand{\rismomb}{{\mbox{\scriptsize RI/SMOM}_{\gamma_{\mu}}}}
\newcommand{\rismom}{\mbox{\scriptsize RI/SMOM}}
\newcommand{\als}{\alpha_s}
\newcommand{\Tr}{\mbox{Tr}}
\newcommand{\Fslash}[1]{\!\not{\hbox{\kern-2pt ${#1}$}}}
\newcommand{\unitop}{1{\hbox{\kern-7pt $1$}}}
\newcommand{\CF}{C_F}
\newcommand{\CA}{C_A}
\newcommand{\TF}{T_F}
\newcommand{\nf}{n_f}
\newcommand{\bA}{\bar{A}}
\newcommand{\bB}{\bar{B}}
\newcommand{\bC}{\bar{C}}
\newcommand{\bD}{\bar{D}}

\begin{document}    

\begin{titlepage}
\noindent
\mbox{}\\[-1cm]
\mbox{}\hfill CU-TP-1186\\
\mbox{}\hfill KANAZAWA-09-01\\
\mbox{}\hfill RBRC-771\\
\mbox{}\hfill SHEP 09/02\\
%

\vspace{0.5cm}
\begin{center}
  \begin{Large}
    \begin{bf}
      \begin{center}
      Renormalization of quark bilinear\\[-0.2cm]
      operators in a momentum-subtraction scheme\\[-0.2cm] 
      with a nonexceptional subtraction point
      \end{center}
    \end{bf}
  \end{Large}
  \vspace{0.8cm}

  \begin{large}
    C. Sturm $\rm^{a \,}$,
    Y. Aoki $\rm^{b \,}$, 
    N.H. Christ $\rm^{c \,}$, 
    T. Izubuchi $\rm^{a,b,d\,}$,\\
    C.T.C. Sachrajda $\rm ^{c,e\,}$
    {\normalsize and } A. Soni $\rm^{a \,}$\\
  \end{large}
    (RBC and UKQCD Collaborations)
  \vskip .7cm
	{\small {\em 
	    $\rm ^a$ 
	     Physics Department, 
             Brookhaven National Laboratory, \\
             Upton,
	     New York 11973, USA}}\\
	{\small {\em 
	    $\rm ^b$
             RIKEN-BNL Research Center,
             Brookhaven National Laboratory, \\
             Upton,
	     New York 11973, USA}}\\
	{\small {\em 
	    $\rm ^c$
                Physics Department, Columbia University, New York, New York
	    10027, USA}}\\
	{\small {\em 
	    $\rm ^d$ 
             Institute for Theoretical Physics,
             Kanazawa University,\\ Kakuma, Kanazawa, 920-1192, Japan}}\\
	{\small {\em 
	    $\rm ^e$ 
            School of Physics and Astronomy, University of Southampton,\\ 
            Southampton SO17 1BJ, United Kingdom}}\\
	\vspace{1cm}
{\bf Abstract}
\end{center}
\begin{quotation}
  \noindent
    We extend the Rome-Southampton regularization independent
    momentum-subtraction renormalization scheme($\mbox{RI/MOM}$) for
    bilinear operators to one with a nonexceptional, symmetric
    subtraction point.  Two-point Green's functions with the insertion
    of quark bilinear operators are computed with scalar, pseudoscalar,
    vector, axial-vector and tensor operators at one-loop order in
    perturbative QCD. We call this new scheme $\mbox{RI/SMOM}$, where
    the S stands for "symmetric". Conversion factors are derived, which
    connect the $\mbox{RI/SMOM}$ scheme and the $\overline{\mbox{MS}}$
    scheme and can be used to convert results obtained in lattice
    calculations into the $\overline{\mbox{MS}}$ scheme. Such a
    symmetric subtraction point involves nonexceptional momenta implying
    a lattice calculation with substantially suppressed contamination
    from infrared effects. Further, we find that the size of the
    one-loop corrections for these infrared improved kinematics is
    substantially decreased in the case of the pseudoscalar and scalar
    operator, suggesting a much better behaved perturbative
    series. Therefore it should allow us to reduce the error in the
    determination of the quark mass appreciably.
\end{quotation}
\end{titlepage}
\section{Introduction}
\label{sec:Introduction}

Lattice simulations in quantum chromodynamics (QCD) allow for ab initio
nonper\-tur\-ba\-tive determinations of operator matrix elements and
physical quantities such as quark masses and the strong coupling
constant. One starts with a direct computation of the bare quantities
with the lattice spacing acting as the ultraviolet cutoff in some
particular discretization of QCD. Providing that the lattice spacing is
sufficiently small, it is in principle possible to obtain the
corresponding renormalized quantities using perturbation
theory. However, the coefficients in lattice perturbation theory
frequently prove to be large and for this reason techniques using
nonperturbative renormalization (NPR) have been developed and are being
successfully implemented. With these techniques lattice perturbation
theory is avoided entirely, and one obtains renormalized quantities in
some appropriate renormalization scheme such as the regularization
independent momentum-subtraction ($\ARI$)
scheme~\cite{Martinelli:1994ty}. 

On the other hand perturbative calculations in continuum QCD are
conventionally and conveniently performed using dimensional
regularization \cite{tHooft:1972fi} and the $\MSbar$ renormalization
scheme~\cite{tHooft:1973mm,Bardeen:1978yd} which is not directly
amenable to the NPR procedure. The continuum perturbation theory is
therefore used to match the quantities computed in the $\ARI$ and
$\MSbar$ schemes. For example the computation of the mass conversion
factor $C^{\ari}_m$, which converts a quark mass renormalized in the
$\ARI$ scheme into the $\MSbar$ scheme or the conversion factor
$C^{\ari}_q$, which performs the corresponding conversion of the quark
fields, are both known up to three-loop order in perturbative
QCD~\cite{Martinelli:1994ty,Franco:1998bm,Chetyrkin:1999pq}.  Another
scheme, which is useful in lattice simulations is the $\RIp$ scheme in
which these conversion factors are also known up to three-loop
order\cite{Chetyrkin:1999pq,Gracey:2003yr}. A more detailed definition
of these schemes will be discussed in
Section~\ref{sec:GeneralNotations}. The conversion factors $C_m$ and
$C_q$ in both schemes can be obtained through the evaluation of
self-energy diagrams. Not only quark masses, but also the strong
coupling constant $\alpha_s$ has been studied in MOM schemes
\cite{Celmaster:1979km,Braaten:1981dv,Jegerlehner:1998zg,Chetyrkin:2000fd,Chetyrkin:2000dq,Chetyrkin:2008jk}.

\begin{figure}[!ht]
\begin{center}
\includegraphics[bb=126 575 262 713,width=3cm]{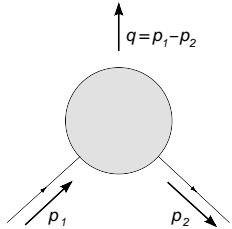}
\end{center}
\caption{\label{fig:mom} Momentum flow of a generic diagram required for
the renormalization procedure with nonexceptional momenta. The gray
bubble stands for an operator insertion and higher order corrections.}
\end{figure}

With regard to the vertex diagrams one has many choices of defining the
subtraction point at which the renormalization constants are fixed
through different momentum configurations. In this paper we determine
the one-loop matching coefficients for a generalization of the
$\ARI$ scheme in which there are no channels with exceptional momenta
and which we propose to use in our numerical simulations. Because the
kinematic configuration in this scheme is symmetrical in the three
channels, we call it the $\RISMOM$ scheme. In the following we define the
symmetric and asymmetric Minkowski momentum configurations by\\[-0.8cm]
\begin{itemize}
\item symmetric or nonexceptional momentum configuration:\\[-0.5cm]
\[
p_1^2=p_2^2=q^2=-\mu^2,\quad \mu^2>0,\quad q=p_1-p_2,
\]\\[-1.4cm]
\item asymmetric or exceptional momentum configuration:\\[-0.5cm]
\[
p_1^2=p_2^2=-\mu^2,\quad \mu^2>0,\quad p_1=p_2,\quad q=0,
\]\\[-1.5cm]
\end{itemize}
where the momentum flow is shown diagrammatically in Fig.~\ref{fig:mom}.

In Ref.~\cite{Allton:2008pn} quark masses were determined through
lattice simulations using nonperturbative
renormalization~\cite{Martinelli:1994ty} in the $\ARI$ scheme and
subsequently converted to the $\MSbar$ scheme. In order to renormalize
the bare quark masses in the lattice simulation, the renormalization
constants need to be computed on the lattice. In regularization and
renormalization schemes which preserve flavor and chiral symmetries in
the limit of vanishing quark masses, the perturbative renormalization
constants of the axial-vector and vector operators as well as the ones
for the pseudoscalar and scalar operators need to be equal. In the
standard $\ARI$ and $\RIp$ schemes the normalization conditions for
quark bilinear operators are imposed on Green's functions with the
operator inserted between equal incoming and outgoing momenta say $p$,
and $-p^2\equiv\mu^2$ is the renormalization scale. The momentum $q$
inserted at the operator is therefore $0$ so that there is an
\textit{exceptional} channel, i.e. one in which the square of the
momentum is much smaller than the typical large scale ($\mu^2$).  For
the asymmetric subtraction point effects of chiral symmetry breaking
vanish only slowly like $1/p^2$ for large external momenta $p^2$. In
Ref.~\cite{Aoki:2007xm} it was proposed instead to use a similar
renormalization procedure but with the incoming and outgoing quarks
having different momenta, $p_1$ and $p_2$ respectively, with
$p_1^2=p^2_2=(p_1-p_2)^2\equiv p^2$. There are now no exceptional
channels and we explain below that this decreases chiral symmetry
breaking and other unwanted infrared effects. The choice of such a
symmetric subtraction point is very convenient, the renormalized
quantities depend also only on a single scale $p^2$.  When the
renormalization constants of quark bilinear operators are fixed at a
symmetric subtraction point (chosen to have nonexceptional kinematics)
chiral symmetry breaking and other unwanted infrared effects are better
behaved and vanish with larger asymptotic powers of the order
$1/p^6$. This behavior has been derived in Ref.~\cite{Aoki:2007xm} as a
consequence of Weinberg's theorem~\cite{Weinberg:1959nj} and
demonstrated by explicitly computing the renormalization constants on
the lattice. Hence these $\RISMOM$ kinematics suppress infrared effects
much more strongly than the usual exceptional configuration for large
external momenta. The symmetric momentum configuration is thus much more
favorable.  However, in order to be able to use it to evaluate the
matrix elements of quark bilinear operators and the quark mass, the
matching factors need to be determined perturbatively for this new,
symmetric choice of momenta. A nonperturbative test of the
$\RISMOM$ scheme for the quark mass renormalization can be found in
Ref.~\cite{Aoki:lat2008}.

Another drawback in the case of the exceptional momenta is that the
perturbative expansion of the usual conversion factor $C_m^{\ari}$ shows
poor convergence and makes a significant contribution to the systematic
uncertainty in the quark masses obtained from the lattice studies. In
fact, in Ref.~\cite{Allton:2008pn} the error ($\approx 11\%$) in the
quark masses arising from the truncation of the perturbative series in
the matching factor amounts to around 60\% of the total error. Therefore
determining the conversion factor for a symmetric momentum configuration
will also allow us to see if the convergence will be better behaved. If
it is better behaved, then the symmetric configuration would be
preferred for both of these reasons.  Motivated by these considerations
we study in this work the renormalization of quark bilinear nonsinglet
operators of the form $\hat{O}=\bar{u}\Gamma d$ for a symmetric
subtraction point, where $\Gamma$ represents a Dirac matrix and
$\bar{u}$ and $d$ are fermion quark fields.

Even with the use of the symmetric, nonexceptional kinematics, the
renormalization prescription is not unique and the chiral
Ward-Takahashi identities can be satisfied using a variety of
procedures. In the following sections we study a specific scheme which
we consider to be convenient and practicable for the nonperturbative
renormalization of lattice quark bilinear operators. In order to
preserve the Ward-Takahashi identity, the definitions of the vertex and
wave function renormalizations are related as we explain in the
following Section.

The outline of this paper is as follows: In
Section~\ref{sec:GeneralNotations} we define our notation and
conventions and introduce the framework required for performing
renormalization of the quark bilinear operators with a symmetric
subtraction point.  Subsequently we present in
Section~\ref{sec:Calculations} two methods for the extraction of the
conversion factor $C_m$ in the $\RISMOM$ scheme, apply the concepts of
Section~\ref{sec:GeneralNotations} to calculate the vector,
axial-vector, pseudoscalar, scalar and tensor operators between two
off-shell quark states at one-loop order in perturbative QCD for the
nonexceptional momentum configuration and determine the matching
factors. Finally we close with a brief summary and our conclusions in
Section~\ref{sec:DiscussConclude}.  Even with the symmetric
nonexceptional kinematics the choice of renormalization conditions is
not unique. In Appendix~\ref{app:AltProj} we therefore present the
one-loop perturbative results in a form which can be used to calculate
the conversion factors from a general scheme with a symmetric
subtraction point to the $\MSbar$ scheme. For illustration we study one
alternative scheme called the $\RISMOMB$ scheme, in which the vertex
renormalization condition is the same as in the $\ARI$ scheme, but with
nonexceptional kinematics and with a different wave function
renormalization. We also provide the results for the conversion factors
and, in Appendix~\ref{app:anomalous}, the corresponding two-loop
anomalous dimensions.
\section{Concepts and framework of the $\RISMOM$ scheme\label{sec:GeneralNotations}}
We will begin with a bare, continuum theory of QCD which has been
regulated using a scheme which guarantees that Green's functions
involving the quark field and quark field bilinears obey the usual
chiral and flavor symmetries of QCD. Dimensional regularization is an
example of such a scheme.

\begin{figure}[!ht]
\begin{center}
\begin{minipage}{2.5cm}
\includegraphics[bb=125 570 248 673,width=2.5cm]{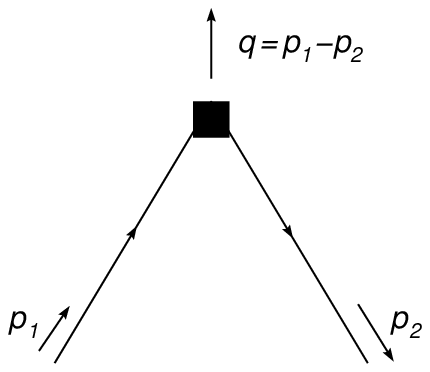}
\vspace{-1cm}
\begin{center}
{\scriptsize{(a)}}
\end{center}
\end{minipage}
\hspace{0.5cm}
\begin{minipage}{2.5cm}
\includegraphics[bb=134 544 278 665,width=2.5cm]{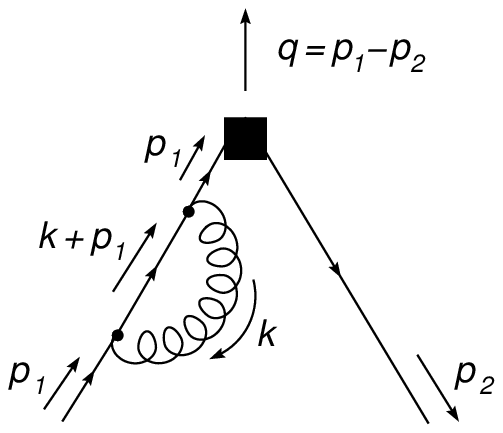}
\vspace{-1cm}
\begin{center}
{\scriptsize{(b)}}
\end{center}
\end{minipage}
\hspace{0.5cm}
\begin{minipage}{2.5cm}
\includegraphics[bb=125 570 248 673,width=2.5cm]{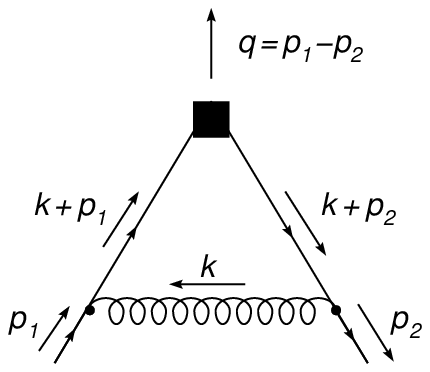}
\vspace{-1cm}
\begin{center}
{\scriptsize{(c)}}
\end{center}
\end{minipage}
\hspace{0.5cm}
\begin{minipage}{2.5cm}
\includegraphics[bb=134 544 278 665,width=2.5cm]{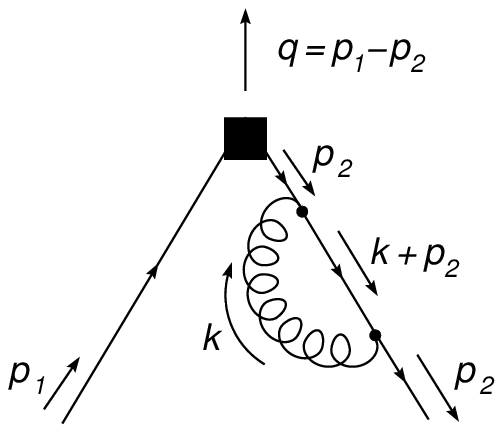}
\vspace{-1cm}
\begin{center}
{\scriptsize{(d)}}
\end{center}
\end{minipage}
\end{center}
\vspace{-0.5cm}
\caption{\label{fig:vertex} Diagrams contributing to the nonamputated
  Green's function up to one-loop order in perturbative QCD. The black
  box indicates the inserted operator. Spiral lines denote gluons and
  solid lines fermions.}
\end{figure}

Let us consider the nonamputated Green's function $G_{\hat{O}}$ of an
operator $\hat{O}$ computed between two external off-shell quark lines
in a fixed gauge. The corresponding diagrams up to one-loop order in
perturbative QCD are shown in Fig.~\ref{fig:vertex}.

\begin{figure}[!ht]
\begin{center}
\begin{minipage}{2.5cm}
\includegraphics[bb=141 617 248 644,width=2.5cm]{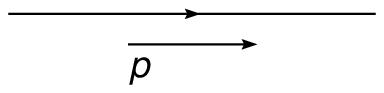}
\vspace{-1cm}
\begin{center}
{\scriptsize{(a)}}
\end{center}
\end{minipage}
\hspace{1cm}
\begin{minipage}[t]{4cm}
\includegraphics[bb=71 618 250 722,width=4cm]{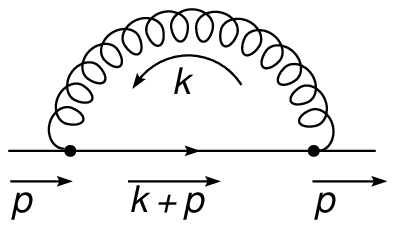}
\vspace{-1cm}
\begin{center}
\hspace{1.35cm}{\scriptsize{(b)}}
\end{center}
\end{minipage}
\end{center}
\vspace{-0.5cm}
\caption{\label{fig:propagator}Propagator-type diagrams up to one-loop
  order in QCD}
\end{figure}

\noindent
The amputated Green's function is defined by
\begin{equation}
\label{eq:amputate}
\Lambda_{\hat{O}}=S^{-1}(p_2)\*G_{\hat{O}}\*S^{-1}(p_1)
\end{equation}
where $S(p)$ is given by the quark propagator
\begin{equation}
-i\*S(p)=\int\!dx e^{ipx}\langle T[\Psi(x)\overline{\Psi}(0)]\rangle=
{i\over \Fslash{p}-m+i\*\ep-\Sigma(p)},
\end{equation}
where $\Sigma(p)$ contains the higher order corrections and can, in
perturbation theory, be decomposed into its Lorentz structure:
$\Sigma(p)=\Fslash{p}\Sigma_V(p^2)+m\*\Sigma_S(p^2)$. The lowest order
and one-loop diagrams contributing to $\Sigma(p)$ are shown in
Fig.~\ref{fig:propagator}.

\noindent
In the following we will consider quark bilinear operators
$\hat{O}=\bar{u}\Gamma d$ with scalar ($\Gamma=\unitop$), pseudoscalar
($\Gamma=i\*\gamma_5$), vector ($\Gamma=\gamma^{\mu}$), axial-vector
($\Gamma=\gamma^{\mu}\gamma_5$) and tensor
($\Gamma=\sigma^{\mu\nu}={i\over2}[\gamma^\mu,\gamma^\nu]$) kernels. We
will distinguish between bare and renormalized quantities by assigning
the index {\scriptsize{$B$}} to a bare quantity and the index
{\scriptsize{$R$}} to a renormalized one. In the case of renormalized
quantities an additional quantifier specifying the scheme is attached.
Renormalized and bare quantities are related through the renormalization
constants~$Z$:
\begin{equation}
\label{eq:RenConst}
\Psi_R=Z_{q}^{1/2}\*\Psi_B,\qquad
m_R=Z_m\*m_B,\qquad
\hat{O}_R=Z_{\hat{O}}\*\hat{O}_B.
\end{equation}
The renormalization constants of the scalar ($\hat{O}=S$), pseudoscalar
($\hat{O}=P$), vector ($\hat{O}=V$), axial-vector ($\hat{O}=A$) and
tensor ($\hat{O}=T$) operator will be denoted as $Z_S$, $Z_P$, $Z_V$,
$Z_A$ and $Z_T$, respectively. In the $\ARI$ scheme the renormalization conditions
which fix the renormalization constants $Z_m$ and $Z_q$ are given by
\begin{equation}
\label{eq:RI}
\lim_{m_R\to0}\left.{1\over12\*m_R}\Tr[S_R^{-1}(p)]\right|_{p^2=-\mu^2}=1
\quad\mbox{and}\quad
\lim_{m_R\to0}\left.{1\over48}\Tr\left[\gamma^{\mu}{\partial S_R^{-1}(p)\over\partial p^{\mu}}\right]\right|_{p^2=-\mu^2}=-1,
\end{equation}
where the symbol ``$\Tr$'' denotes the trace over color and spins.  The
second equation determines $Z^{\ari}_q$ and subsequently the first one
can be used to extract $Z^{\ari}_m$. Now in the $\RIp$ scheme the second
condition of Eqs.(\ref{eq:RI}) is replaced by
\begin{equation}
\label{eq:RIp}
\lim_{m_R\to0}\left.{1\over12\*p^2}\Tr[S_R^{-1}(p)\*\Fslash{p}]\right|_{p^2\to-\mu^2}=-1.
\end{equation}
The quark propagator in the $\RIp$ scheme is fixed to its lowest order
value at the point $p^2=-\mu^2$, where $p^2$ is the squared, external,
Minkowski momentum and $\mu$ is the renormalization scale.

The propagator and vertex diagrams (Fig. \ref{fig:vertex}) for the
vector and axial-vector operators are related through the vector
Ward-Takahashi identity for degenerate masses $m_u=m_d=m$
\begin{equation}
\label{eq:vectorWI}
q_\mu\Lambda^{\mu}_{V,B}(p_1,p_2)=S_{B}^{-1}(p_2)-S_{B}^{-1}(p_1)
\end{equation}
and the axial-vector Ward-Takahashi identity
\begin{equation}
\label{eq:axialWI}
-i\*q_\mu\Lambda^{\mu}_{A,B}(p_1,p_2)=2\*m_B\*\Lambda_{P,B}(p_1,p_2)
       -i\*\gamma_5\*S_{B}^{-1}(p_1)-S_{B}^{-1}(p_2)\*i\*\gamma_5,
\end{equation}
with the momentum transfer $q=p_1-p_2$. The renormalized and bare
amputated Green's functions are connected by
\begin{equation}
\label{eq:AmpGreenRen}
S_{R}(p)=Z_q\*S_B(p),
\qquad
\Lambda_{\hat{O},R}(p_1,p_2)={Z_{\hat{O}}\over Z_q}\Lambda_{\hat{O},B}(p_1,p_2).
\end{equation}
In the following we want to renormalize the quark bilinear operators
using a symmetric subtraction point. For functions $f$, which are
restricted to the symmetric momentum configuration we use the shorthand
$f(p_1^2,p_2^2,q^2)|_{p_1^2=p_2^2=q^2=-\mu^2} \equiv
f(p_1^2,p_2^2,q^2)|_{sym}$ and for the asymmetric subtraction point we
introduce the abbreviation $f(p_1^2,p_2^2,q^2)|_{q=0,
p_1^2=-\mu^2=p_2^2} \equiv f(p_1^2,p_2^2,q^2)|_{asym}$.

We perform the quark mass and wave function renormalization by imposing
on the two-point function $S(p)$ the condition of Eq.(\ref{eq:RIp}) and
\begin{equation}
\label{eq:sRIS}
\lim_{m_R\to0}
{1\over12\*m_R}\left\{
\left.\Tr\left[S^{-1}_R(p)\right]\right|_{p^2=-\mu^2}
-\left.{1\over2}\*\Tr\left[q_{\mu}\Lambda^{\mu}_{A,R}(p_1,p_2)\gamma_5\right]\right|_{sym}
\right\}
=1.
\end{equation}
The second term in the curly brackets on the left-hand side of
Eq.(\ref{eq:sRIS}) starts at $\mathcal{O}(\alpha_s)$ and
is absent in the $\ARI$ and $\RIp$ schemes. This term is needed to
maintain the Ward-Takahashi identities for renormalized quantities, as
we will see below.  For the vector and axial-vector quark bilinear
operators we impose the conditions
\begin{equation}
\label{eq:sRIVA}
\lim_{m_R\to0}\left.{1\over12\*q^2}\Tr\left[
q_{\mu}\Lambda^{\mu}_{V,R}(p_1,p_2)\*\Fslash{q}
\right]\right|_{sym}\!\!\!\!\!=1,\qquad
\lim_{m_R\to0}\left.{1\over12\*q^2}\Tr\left[
q_{\mu}\Lambda^{\mu}_{A,R}(p_1,p_2)\*\gamma_5\*\Fslash{q}
\right]\right|_{sym}\!\!\!\!\!\!=1.
\end{equation}
The projectors for the amputated Green's functions in
Eqs.(\ref{eq:sRIVA}) are different from those used in the
$\ARI$ scheme(see Table~\ref{tab:Cond}). Using instead these original
$\ARI$ projectors leads to a different wave function renormalization and
will be discussed in Appendix~\ref{app:AltProj}. For the pseudoscalar
and scalar amputated Green's functions we use the renormalization
conditions
\begin{equation}
\label{eq:sRIPS}
\lim_{m_R\to0}\left.{1\over12\*i}\Tr\left[\Lambda_{P,R}(p_1,p_2)\gamma_5\right]\right|_{sym}\!\!\!\!\!\!=1,\qquad\quad
\lim_{m_R\to0}\left.{1\over12}\Tr\left[\Lambda_{S,R}(p_1,p_2)\unitop\right]\right|_{sym}\!\!\!\!\!\!=1,
\end{equation}
and for the tensor operator the condition
\begin{equation}
\label{eq:sRIT}
\lim_{m_R\to0}\left.{1\over 144}\Tr\left[\Lambda^{\mu\nu}_{T,R}\sigma_{\mu\nu}\right]\right|_{sym}=1.
\end{equation}

Note that all of the renormalization schemes being considered in this
paper are mass-independent.  Thus, each condition is imposed at fixed
external momentum and vanishing quark mass.  The renormalization
conditions of the $\ARI$ and $\RISMOM$ schemes are summarized
in Table~\ref{tab:Cond}.
\begin{table}[!ht]
\begin{tabular}{|l|l|}
\hline
$\ARI$&
$
\lim\limits_{m_R\to0}\left.{1\over48}\Tr\left[\gamma^{\mu}{\partial
    S_R^{-1}(p)\over\partial p^{\mu}}\right]\right|_{p^2=-\mu^2}=-1,
\quad
\lim\limits_{m_R\to0}\left.{1\over12\*m_R}\Tr[S_R^{-1}(p)]\right|_{p^2=-\mu^2}=1
$,\\&
$
\lim\limits_{m_R\to0}\left.{1\over48}\Tr\left[
\Lambda^{\mu}_{V,R}(p_1,p_2)\*\gamma_{\mu}
\right]\right|_{asym}\!\!\!\!\!=1,\quad
\lim\limits_{m_R\to0}\left.{1\over48}\Tr\left[
\Lambda^{\mu}_{A,R}(p_1,p_2)\*\gamma_5\*\gamma_{\mu}
\right]\right|_{asym}\!\!\!\!\!\!=1
$,\\&
$
\lim\limits_{m_R\to0}\left.{1\over12}\Tr\left[\Lambda_{S,R}(p_1,p_2)\unitop\right]\right|_{asym}\!\!\!\!\!\!=1,\quad
\lim\limits_{m_R\to0}\left.{1\over12\*i}\Tr\left[\Lambda_{P,R}(p_1,p_2)\gamma_5\right]\right|_{asym}\!\!\!\!\!\!=1
$.
\\ \hline
$\RISMOM$&
$
\lim\limits_{m_R\to0}\left.{1\over12\*p^2}\Tr[S_R^{-1}(p)\*\Fslash{p}]\right|_{p^2=-\mu^2}=-1
$,\\&
$
\lim\limits_{m_R\to0}{1\over12\*m_R}\*\left\{
\left.\Tr\left[S^{-1}_R(p)\right]\right|_{p^2=-\mu^2}
-{1\over2}\*\left.\Tr\left[q_{\mu}\Lambda^{\mu}_{A,R}(p_1,p_2)\gamma_5\right]\right|_{sym}
\right\}=1
$,\\&
$
\lim\limits_{m_R\to0}\left.{1\over12\*q^2}\Tr\left[
q_{\mu}\Lambda^{\mu}_{V,R}(p_1,p_2)\*\Fslash{q}
\right]\right|_{sym}\!\!\!\!\!=1,\quad
\lim\limits_{m_R\to0}\left.{1\over12\*q^2}\Tr\left[
q_{\mu}\Lambda^{\mu}_{A,R}(p_1,p_2)\*\gamma_5\*\Fslash{q}
\right]\right|_{sym}\!\!\!\!\!\!=1
$,\\&
$
\lim\limits_{m_R\to0}\left.{1\over12}\Tr\left[\Lambda_{S,R}(p_1,p_2)\unitop\right]\right|_{sym}\!\!\!\!\!\!=1,\quad
\lim\limits_{m_R\to0}\left.{1\over12\*i}\Tr\left[\Lambda_{P,R}(p_1,p_2)\gamma_5\right]\right|_{sym}\!\!\!\!\!\!=1
$.
\\ \hline
\end{tabular}
\caption{\label{tab:Cond} The renormalization conditions for the
  $\ARI$ and $\RISMOM$ schemes. }
\end{table}

In the remainder of this Section we will show that if the normalization
conditions in Eqs.(\ref{eq:RIp}) and (\ref{eq:sRIS})-(\ref{eq:sRIPS}) of
this $\RISMOM$ scheme are imposed on the quark bilinear operators, the
Ward-Takahashi identities of Eqs.(\ref{eq:vectorWI}) and (\ref{eq:axialWI})
are also obeyed for the resulting renormalized quantities and the
properties $Z_V=1=Z_A$, $Z_P=1/Z_m$ and $Z_S=Z_P$ are preserved, as they
are in the $\MSbar$, $\RIp$ and $\ARI$ schemes (see
e.g.~Ref.\cite{Chetyrkin:1996ia,Blum:2001sr}). Some of
these properties hold nonperturbatively while the others are proven
only in the perturbation theory as we will see below.\\

Let us start by considering the object ${1\over
12q^2}\Tr[q_{\mu}\Lambda_{V,B}^{\mu}\*\Fslash{q}]|_{sym}$ and insert
the vector Ward-Takahashi identity of Eq.(\ref{eq:vectorWI}):
\begin{equation}
\label{eq:ZqRIpRel}
\!\!\!\!\!\left.{1\over 12q^2}\!\Tr[q_{\mu}\Lambda_{V,B}^{\mu}\Fslash{q}]
\right|_{sym}\!\!\!\!\!\!\!=
{1\over 12q^2}\!\!\left.\left\{\!\Tr[S_{B}^{-1}\!(p_2)\Fslash{q}]
\!-\!\Tr[S_{B}^{-1}\!(p_1)\Fslash{q}]\!\right\}\right|_{sym}\!\!\!\!\!\!\!=
\!-{1\over 12q^2}\!\!\left.\Tr[S_{B}^{-1}\!(q)\Fslash{q}]\right|_{sym}.\!\!\!\!\!\!\!\!\!\!
\end{equation}
Expressing bare quantities in terms of renormalized ones using
Eq.(\ref{eq:AmpGreenRen}) and imposing the condition in
Eq.(\ref{eq:RIp}) and the one on the left in Eq.(\ref{eq:sRIVA}) leads
to $Z^{\rismom}_V=1$. Similarly one obtains
$Z^{\rismom}_V=Z^{\rismom}_A$ by inserting Eq.(\ref{eq:axialWI}) into
${1\over
12q^2}\Tr[q_{\mu}\Lambda_{A,B}^{\mu}\*\gamma_5\*\Fslash{q}]|_{sym}$,
combining it with Eqs.(\ref{eq:ZqRIpRel}) and imposing the conditions of
Eq.(\ref{eq:sRIVA}) for the renormalized quantities in the massless
limit. Note that the above derivation of $Z_A^{\rismom}=Z_V^{\rismom}$
is independent of the choice of the renormalization point $\mu$. This is
in contrast to the $\ARI$ scheme for which the Ward-Takahashi identity
for the axial current only holds at large $\mu^2$. The renormalized
vector current satisfies the Ward-Takahashi identity in both the $\ARI$
and the $\RISMOM$ schemes even in the low energy region. However, the
relation $Z_A = Z_V =1$ implies that the axial vertex function given in
Eq.~(\ref{eq:sRIVA}) remains exactly equal to one in the limit of
vanishing quark mass even when evaluated in the infrared region of QCD
where large vacuum chiral symmetry breaking might have been expected to
introduce large asymmetries between such vector and axial-vector
correlation functions.

From $Z^{\rismom}_V=1=Z^{\rismom}_A$ it follows that the renormalization
constant $Z_q^{\rismom}$ can be extracted from Eqs.(\ref{eq:sRIVA}).
However, since Eq.(\ref{eq:RIp}) determines $Z_q$ in both the $\RISMOM$
and $\RIp$ schemes, $Z_q^{\rismom}=Z_q^{\rip}$, whose value is known up
to order $\alpha_s^3$ in
Ref.~\cite{Chetyrkin:1999pq,Gracey:2003yr}. Nevertheless, in
Section~\ref{sec:vectoraxial}, we will renormalize the vector and
axial-vector operators for the symmetric momentum configuration in the
$\RISMOM$ scheme using the conditions in Eqs.(\ref{eq:sRIVA}) in order
to demonstrate that the value of $Z_q^{\rismom}$ obtained from
Eq.(\ref{eq:sRIVA}) is in fact equal to the value for $Z_q^{\rip}$
obtained from Eq.(\ref{eq:RIp}) by explicit calculation up to one-loop
order.\\

From the axial Ward-Takahashi identity it follows that the
renormalization constant for the pseudoscalar operator $Z^{\rismom}_P$
and the mass renormalization constant $Z^{\rismom}_m$ are related. If
one multiplies Eq.(\ref{eq:axialWI}) by $(-i\gamma_5)$, takes the trace
of both sides over spin and color and restricts it to the symmetric
momentum configuration, one obtains
\begin{equation}
\label{eq:SimpProjAxialWI}
-\left.{1\over12}\Tr[q_\mu\Lambda_{A,B}^{\mu}\*\gamma_5]\right|_{sym}=
2\*m_B\*\left.{1\over12\*i}\Tr[\Lambda_{P,B}\*\gamma_5]\right|_{sym}
-{1\over6}\*\left.\Tr[S_B(p)^{-1}]\right|_{p^2=-\mu^2}.
\end{equation}
Taking the zero-mass limit, expressing again the bare equation with the
help of Eqs.(\ref{eq:RenConst}) and (\ref{eq:AmpGreenRen}) in terms
of renormalized quantities and imposing the conditions in
Eqs.(\ref{eq:sRIS}) and (\ref{eq:sRIPS}) for the $\RISMOM$ scheme leads
to $Z^{\rismom}_P=1/Z^{\rismom}_m$.\\

The conditions in Eq.(\ref{eq:sRIPS}) for the pseudoscalar and scalar
operator can be expressed in terms of the bare Green's function and the
renormalization constants. The traces over the two bare Green's functions
become equal in the massless limit in perturbation theory, which leads 
to $Z_S=Z_P$.\\

In the above discussion the renormalization constants relate the
bilinear operators renormalized in the $\RISMOM$ scheme to those in the
bare theory which we had assumed to satisfy the
Ward-Takahashi identities (\ref{eq:vectorWI}) and
(\ref{eq:axialWI}). Since many lattice formulations of QCD break the
chiral or flavor symmetries, in general Eqs.(\ref{eq:vectorWI}) and
(\ref{eq:axialWI}) do not hold in these (bare) theories. Nevertheless,
our renormalization scheme is indeed \textit{regularization independent}
and the Ward-Takahashi identities hold for the $\RISMOM$ renormalized
quantities. The renormalization constants relating the renormalized and
bare lattice operators depend on the regularization of course, so that,
for example, $Z_V$ and $Z_A$ will typically be different from $1$ in
such cases. 

%
%
%
%
%
%
\section{Conversion factors:\\Results of the next-to-leading order
  calculation\label{sec:Calculations}} 
The properties discussed in Section~\ref{sec:GeneralNotations} can be
used to convert quark masses determined through lattice simulations in
the $\RISMOM$ scheme into the $\MSbar$ scheme by computing the matching
factor $C^{\rismom}_m={Z^{\msbar}_m/ Z^{\rismom}_m}$ with
$m^{\msbar}_R=C^{\rismom}_m\*m^{\rismom}_R$. The explicit calculation to
determine this conversion factor at one-loop order in perturbative QCD
will be performed in the next subsections using two different methods,
which allows us to cross-check our results.\\

First, the matching factor $C^{\rismom}_m$ can be obtained with the help
of Eq.(\ref{eq:sRIS}) through
\begin{equation}
\label{eq:CmsRISelf}
(C^{\rismom}_m)^{-1}=(C^{\rip}_m)^{-1}-
      {1\over2}\*C^{\rismom}_q\*\lim_{m_R\to0}
      \left.{1\over12\*m^{\msbar}_R}\*
      \Tr\left[q_{\mu}\Lambda^{\mu,\msbar}_{A,R}\gamma_5\right]\right|_{sym},
\end{equation}
which will be evaluated in Section~\ref{sec:vectoraxial}. In analogy to
$C^{\rismom}_m$ we define here the conversion factor
$C^{\rismom}_q={Z^{\msbar}_{q}}/{Z^{\rismom}_q}$ for the fermion fields.\\

Second, the conversion factor can be related to the renormalization
constants of the pseudoscalar operator
\begin{equation}
\label{eq:CmEq1oCp}
C^{\rismom}_m={Z^{\msbar}_m\over Z^{\rismom}_m}={Z^{\rismom}_P\over Z^{\msbar}_P}\equiv{1\over C^{\rismom}_P}
\end{equation}
and hence
\begin{equation}
\label{eq:conv1}
m^{\msbar}_{R}={1\over C^{\rismom}_P}\*m^{\rismom}_{R}
={1\over C^{\rismom}_P}\* {1\over Z^{\rismom}_{P,latt.}}\*m_{B,latt.}.
\end{equation}
In particular in Section~\ref{sec:pseudoscalar} we will evaluate the
conversion factor $C^{\rismom}_P$, which converts the pseudoscalar
operator from the $\RISMOM$ scheme to the $\MSbar$ scheme. The matching
factor $C^{\rismom}_P$ is in general gauge dependent; however, this
gauge dependence will cancel out with the corresponding gauge dependence
in the factor $Z^{\rismom}_{P,latt.}$ determined in the lattice
calculation. In the following we will perform the computation in the general
covariant gauge using the tree level gluon propagator
\begin{equation}
\label{eq:GluonPropagator}
{i\delta^{ab}\over q^2+i\*\ep}\*\left(-g^{\mu\nu}+(1-\xi)\*{q^{\mu}\*q^{\nu}\over q^2+i\*\ep}\right)
\end{equation}
and we will restrict ourselves to the Landau gauge ($\xi=0$) at the end of
the calculation. We choose the renormalization scales of both schemes to
be equal $\mu^{\msbar}=\mu^{\rismom}$. The conversion factors
$C^{\rismom}_{x}$ with $x\in\{m,q,S,P,V,A,T\}$ denote always the
conversion from the $\RISMOM$ to the $\MSbar$ scheme.
\subsection{The vector and axial-vector operator\label{sec:vectoraxial}}
In this section we want to use the vector and axial-vector operator
separately to extract the matching factor $C^{\rismom}_q$ for the quark
field for the symmetric subtraction point. This result is then used in
the next step to compute $C_m^{\rismom}$ with the help of
Eq.(\ref{eq:CmsRISelf}).

Our perturbative computation is performed in dimensional regularization
with the space-time dimension $d=4-2\*\vep$. For the vector operator
case $C^{\rismom}_q$ can be obtained by
\begin{equation}
\label{eq:CqVector}
(C^{\rismom}_q)^{-1}=\lim_{m_R\to0}\left.{1\over12q^2}\*
\Tr\left[q_{\mu}\Lambda^{\mu,\msbar}_{V,R}\Fslash{q}\right]\right|_{sym}.
\end{equation}
The calculation of the one-loop QCD corrections to the vector operator,
computed between two off-shell quark lines, is straightforward and
leads to
\begin{equation}
\label{eq:CqsRI}
C^{\rismom}_q=1-{\alpha_s\over4\*\pi}\*\CF\*\xi+\mathcal{O}(\alpha_s^2).
\end{equation}
The symbol $\CF$ denotes the Casimir operator of the SU(3) group in the
fundamental representation; $\CF=4/3$. As expected Eq.(\ref{eq:CqsRI})
agrees with the result in Ref.~\cite{Chetyrkin:1999pq,Gracey:2003yr},
since $C^{\rismom}_q=C^{\rip}_q$, as shown in
Section~\ref{sec:GeneralNotations}. 

Similarly one can also derive this result from the axial-vector operator
by using
\begin{equation}
\label{eq:CqAxialVector}
(C^{\rismom}_q)^{-1}=\lim_{m_R\to0}\left.{1\over12q^2}\*
\Tr\left[q_{\mu}\Lambda^{\mu,\msbar}_{A,R}\gamma_5\Fslash{q}\right]\right|_{sym}.
\end{equation}
For the treatment of $\gamma_5$ in dimensional
regularization~\cite{tHooft:1972fi,Breitenlohner:1977hr} we use a naive
anticommuting definition of $\gamma_5$ for evaluating the
loop integrals, which obeys the equations $\{\gamma_5,\gamma^\mu\}=0$
and $\gamma_5^2=1$. This is a self-consistent prescription for the
flavor nonsinglet contributions considered in this
work~\cite{Trueman:1979en,Larin:1993tq}. On the other hand one can use
Eq.(\ref{eq:RIp}) in order to determine $Z^{\rismom}_q$ and then extract
$Z^{\rismom}_V$ and $Z^{\rismom}_A$ from Eqs.(\ref{eq:sRIVA}). For both
we explicitly confirm that at one-loop order
$Z^{\rismom}_V=Z^{\rismom}_A=1$ as expected.

The conversion factor $C^{\rismom}_m$ can now be computed from the
axial-vector operator with the help of Eq.(\ref{eq:CmsRISelf}) by
determining
\begin{equation}
\lim_{m_R\to0} \left.{1\over12\*m^{\msbar}_R}\*
\Tr\left[q_{\mu}\Lambda^{\mu,\msbar}_{A,R}\gamma_5\right]\right|_{sym}=
{\alpha_s\over4\*\pi}\CF\left(3+\xi\right)\*C_0
+\mathcal{O}(\alpha_s^2),
\end{equation}
with 
\begin{equation}
\label{eq:C0}
C_0={2\over3}\Psi'\!\left({1\over3}\right)-\left({2\over3}\*\pi\right)^2,
\end{equation}
where $\Psi(x)$ is the digamma function $\Psi(x)=\Gamma'(x)/\Gamma(x)$
\footnote{The prime denotes here the derivative.}.  The matching factor
$C^{\rip}_m$ can be taken from
Ref.~\cite{Chetyrkin:1999pq,Gracey:2003yr} and $C^{\rismom}_q$ from
Eq.(\ref{eq:CqsRI}).  This leads to
\begin{equation}
\label{eq:Csrim}
C^{\rismom}_m=1-{\alpha_s\over4\*\pi}\*\CF\left(
4+\xi-(3+\xi)\*{1\over2}\*C_0
\right)+\mathcal{O}(\alpha_s^2).
\end{equation}
\subsection{The pseudoscalar and scalar operator\label{sec:pseudoscalar}}
In this section we determine the conversion factor
$C^{\rismom}_m=(C^{\rismom}_P)^{-1}=Z^{\rismom}_P/Z^{\msbar}_P$ through
the calculation of the pseudoscalar operator.  At one-loop order in
perturbative QCD its computation leads to the decomposition
\begin{equation}
\label{eq:PseudoDecomp}
\Lambda_{P,B}=A_{P,B}\*i\*\gamma_5
       +B_{P,B}\*i\*\gamma_5\*{m\*\Fslash{q}\over q^2}
       +C_{P,B}\*i\*\gamma_5\*{[\Fslash{p}_1,\Fslash{p}_2]\over q^2},
\end{equation}
with
\begin{equation}
A_{P,B}=1+{\als\over4\*\pi}\*a_{p,1}+\dots,\quad
B_{P,B}={\als\over4\*\pi}\*b_{p,1}+\dots,\quad
C_{P,B}={\als\over4\*\pi}\*c_{p,1}+\dots,
\end{equation}
where the dots stand for higher order corrections and where we have set
$p_1^2=q^2=p_2^2$. The quantities $b_{p,1}$ and $c_{p,1}$ are
finite, whereas $a_{p,1}$ contains $1/\vep$-poles. In the limit of
massless fermions, considered here, we obtain $b_{p,1}=0$. The matching
factor $C^{\rismom}_P$ can be obtained from Eq.(\ref{eq:sRIPS}) by
evaluating
\begin{equation}
\label{eq:CP}
C^{\rismom}_P=C^{\rismom}_q\lim_{m_R\to0}\left.{1\over12i}\Tr\left[\Lambda^{\msbar}_{P,R}\gamma_5\right]\right|_{sym}.
\end{equation}
The fermion field conversion factor $C^{\rismom}_q$ is known and has
been discussed in the previous Section~\ref{sec:vectoraxial}.  Since the
amplitudes $B_{P,B}$ and $C_{P,B}$ in Eq.(\ref{eq:PseudoDecomp}) do not
contribute to the trace in Eq.(\ref{eq:sRIPS}), this condition depends
only on the pseudoscalar amplitude $A_{P,B}$, which is fixed to its
lowest order value at the symmetric subtraction point.  The explicit
calculation yields
\begin{equation}
\label{eq:conv2}
C^{\rismom}_P=C^{\rismom}_q\*\!\Bigg\{1+{\als\over4\*\pi}\*\CF\*\!
\Bigg[
4\!+\!2\*\xi\!+\*\left(1\!+\!{\xi\over3}\right)\*\!
\left({2\over3}\*\pi^2-\Psi'\!\left({1\over3}\right)\right)
\Bigg]\!+\!\mathcal{O}\left(\alpha^2_s\right)\Bigg\}.\!\!
\end{equation}
Inserting $C^{\rismom}_q$ from Eq.(\ref{eq:CqsRI}) and exploiting
$C^{\rismom}_m=(C^{\rismom}_P)^{-1}$ leads to the same result as given in
Eq.(\ref{eq:Csrim}).  Numerical evaluation in the Landau gauge leads to
\begin{equation}
\label{eq:num}
C^{\rismom}_P=(C^{\rismom}_m)^{-1}=1+{\als\over4\*\pi}\*\CF\*\;0.4841391...+\mathcal{O}\left(\alpha_s^2\right).
\end{equation}
Comparing Eq.(\ref{eq:num}) to the $\RIp$ scheme with
$C^{\rip}_m=1-{\alpha_s\over4\*\pi}\*\CF\*4+\dots$ or the $\ARI$ scheme, 
(which is in the Landau gauge at one-loop order equal to the $\RIp$
scheme),  
we see that the result in Eq.(\ref{eq:num}) has a smaller one-loop
coefficient by almost a factor of 10. \\

\begin{figure}[!ht]
\begin{minipage}{8cm}
\begin{center}
\includegraphics[bb=0 0 567 384,width=8cm]{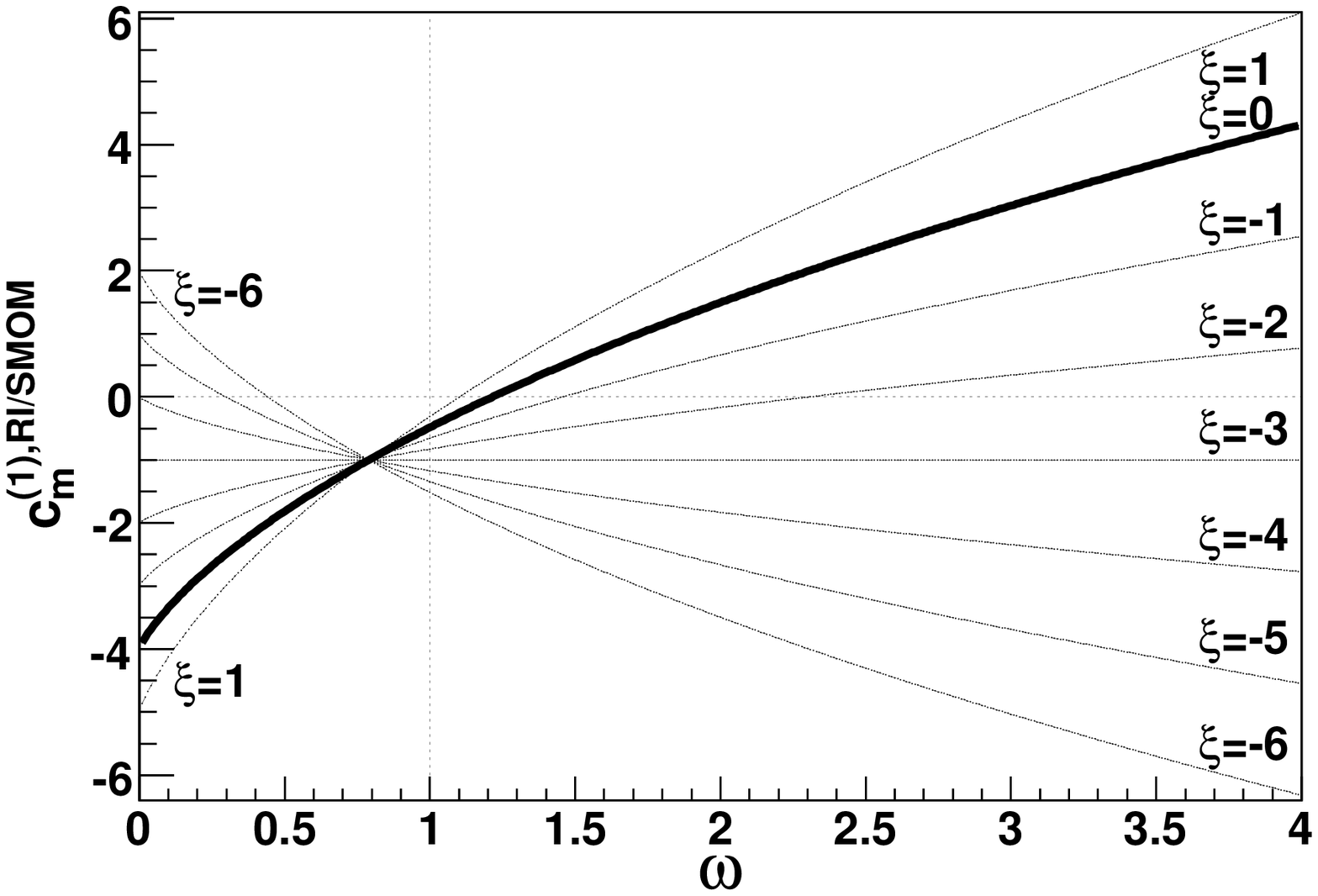}
(a)
\end{center}
\end{minipage}
\begin{minipage}{8cm}
\begin{center}
\includegraphics[bb=0 0 567 384,width=8cm]{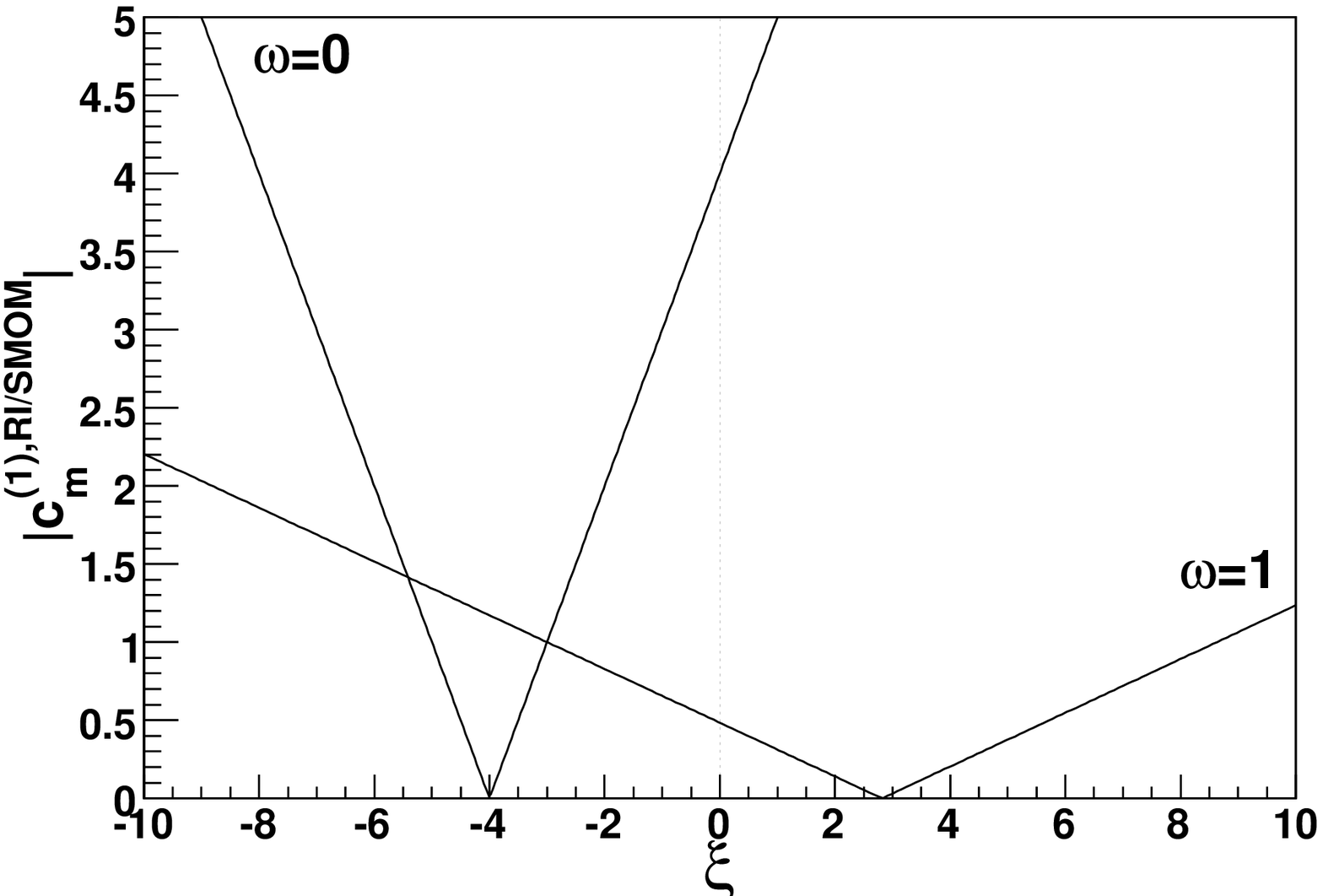}
(b)
\end{center}
\end{minipage}
\caption{\label{fig:cm} (a) shows the one-loop coefficient
  $c^{(1),\rismom}_m$ of the matching factor $C^{\rismom}_m$ as a
  function of $\omega$ for different gauges. The value at $\omega=0$ is
  the result for the exceptional momentum configuration. The
  nonexceptional configuration is indicated through the vertical line
  at $\omega=1$. The bold line indicates the Landau gauge ($\xi=0$), which
  is usually adopted in lattice calculations. (b) shows the
  exceptional ($\omega=0$) and nonexceptional ($\omega=1$)
  configuration as a function of the gauge parameter $\xi$.}
\end{figure}

In order to study the conversion factor $C^{\rismom}_m$ for different
subtraction points, we introduce the parameter $\omega$ and fix our
renormalization condition for the subtraction ``point''
$p_1^2=p_2^2=-\mu^2$ and $q^2=-\omega\*\mu^2$.  This allows us also to
study the limit $\omega\to0$, which results in an exceptional momentum
configuration, whereas the limit $\omega\to1$ gives the symmetric
one. The result depending on $\omega$ is given by
\begin{equation} 
\label{eq:gen}
C^{\rismom}_m=1\!-\!{\als\over4\*\pi}\*\CF\*\!
\left[
4\!+\!\xi\!-\left(3+\xi\right)\*{\omega\over2}\*C_0(\omega)\right]
+\mathcal{O}\left(\alpha^{2}_s\right),
\end{equation}
where the function $C_0(\omega)$ for $\omega\in[0,4]$ is given by
\begin{eqnarray}
C_0(\omega)&=&-\mu^2\left.\int\!{d^4k\over
  i\*\pi^2}{1\over(k+p_1)^2\*(k+p_2)^2\*k^2}\right|_{p_1^2=p_2^2=-\mu^2,
  q^2=-\omega\mu^2}
\nonumber\\
&=&{2\*i\over \sqrt{4-\omega}\*\sqrt{\omega}}
\left[
 \mbox{Li}_2\left({-\sqrt{4-\omega}+i\*\sqrt{\omega}\over
                   -\sqrt{4-\omega}-i\*\sqrt{\omega}}\right)
-\mbox{Li}_2\left({-\sqrt{4-\omega}-i\*\sqrt{\omega}\over
                   -\sqrt{4-\omega}+i\*\sqrt{\omega}}\right)
\right]\!\!,
\end{eqnarray}
and $\mbox{Li}_2(z)$ is the dilogarithm function. In the case $\omega=1$
one obtains the result of Eq.(\ref{eq:Csrim}) with $C_0(\omega=1)=C_0$.
In order to display the dependence of this result for $C^{\rismom}_m$ on
the gauge parameter $\xi$, we introduce the one-loop coefficient
function $c^{(1),\rismom}_m(\omega,\xi)$ extracted from
Eq.(\ref{eq:gen}) using the definition:
$C^{\rismom}_m=1+{\als\over4\*\pi}\*\CF\*c^{(1),\rismom}_m(\omega,\xi)$.
The coefficient $c^{(1),\rismom}_m(\omega,\xi)$ is plotted as a function
of $\omega$ in the interval $\omega\in[0,4]$ for different gauges in
Fig.~\ref{fig:cm}(a).

Going from the exceptional ($\omega=0$) to the nonexceptional
($\omega=1$) momentum configuration leads to a smaller one-loop
coefficient in the Landau gauge; however even for almost all other gauges
the one-loop coefficient becomes smaller as well, except for gauges in
the small interval $\xi\in(\frac{4}{C_0-4}-3,-3)$, which is shown in
Fig.~\ref{fig:cm}(b). The smaller coefficient might indicate that the
symmetric configuration is less disposed to infrared effects. \\

In analogy to the pseudoscalar operator the computation of the scalar
operator leads to the conversion factor $C^{\rismom}_S$ by employing the
renormalization condition in Eq.(\ref{eq:sRIPS}). As expected the
one-loop result for the matching factor
$C^{\rismom}_m=(C^{\rismom}_S)^{-1}$, extracted from the scalar
operator, is equal to the result obtained from the pseudoscalar one.
\subsection{The tensor operator\label{sec:tensor}}
The matching factor converting the Green's function of the tensor
operator from the $\RISMOM$ to the $\MSbar$ scheme can be obtained from
Eq.(\ref{eq:sRIT}) in complete analogy to the other operators. It is
given by
\begin{equation}
\label{eq:CT}
C^{\rismom}_T=1+{\alpha_s\over4\pi}\CF\left[\left(1-\xi\right)\left({C_0\over2}-{4\over3}\right)-\xi\right],
\end{equation}
where we have defined $C^{\rismom}_{T}={Z^{\msbar}_{T} /
Z^{\rismom}_T}$. The numerical evaluation in the Landau gauge leads to
\begin{equation}
C^{\rismom}_T=1-{\alpha_s\over4\pi}\CF\;0.1613797....
\end{equation}
For the $\RIp$ scheme the tensor operator has been evaluated up to
three-loop order in Ref.~\cite{Gracey:2003yr}.  The conversion factor at
one-loop order is found to be proportional to the gauge parameters
$\xi$. This contribution is therefore zero in the Landau gauge.

All conversion factors discussed in Section~\ref{sec:Calculations}
are summarized in Table~\ref{tab:ConvFac}. The matching factors for the
scalar and pseudoscalar operator are equal to the inverse of the mass
conversion factor $1/C^{\rismom}_m=C^{\rismom}_S=C^{\rismom}_P$.
\begin{table}[!h]
\begin{center}
\begin{tabular}{|l|}
\hline
$C^{\rismom}_m=
1-{\als\over4\*\pi}\*\CF\*\;0.4841391...
+\mathcal{O}\left(\alpha_s^2\right)$\\[0.1cm]\hline
$C^{\rismom}_q=
1
+\mathcal{O}\left(\alpha_s^2\right)$\\[0.1cm]\hline
$C^{\rismom}_T=
1-{\alpha_s\over4\pi}\CF\;0.1613797...
+\mathcal{O}\left(\alpha_s^2\right)$\\[0.1cm]\hline
\end{tabular}
\caption{\label{tab:ConvFac} Summary of the matching factors for the
  mass and fermion field conversion $\left(C^{\rismom}_m\right.$,
  $\left.C^{\rismom}_q\right)$ as well as for the conversion of the tensor
  operator $\left(C^{\rismom}_T\right)$ in the Landau gauge. }
\end{center}
\end{table}
\section{Summary and Conclusion\label{sec:DiscussConclude}}
We provide the framework and concepts for renormalizing the quark
bilinear operators in a MOM scheme ($\RISMOM$) with a symmetric
subtraction point which has no channels with exceptional momenta.  This
generally suppresses the infrared chiral symmetry breaking effects
compared to the standard $\ARI$ (or $\RIp$)scheme in which there is an
exceptional channel (with zero momentum). An exception is the vector
current for which the $\ARI$ scheme satisfies the
Ward-Takahashi identity also at low values of $p^2$. We demonstrate that
the chiral Ward-Takahashi identites(for degenerate masses) are satisfied
nonperturbatively, and thus $Z_V=1=Z_A$ for all values of $p^2$, in the
$\RISMOM$ scheme. We calculate the matching factors relating operators
renormalized in this scheme and the $\MSbar$ schemes at
one-loop order in perturbation theory. The one-loop coefficients are
given in Table~\ref{tab:ConvFac} and we note that they are small. In
particular, for the quark mass the coefficient is much smaller than that
between the $\ARI$($\RIp$) and $\MSbar$ schemes which, if confirmed at
higher orders, would lead to a significant reduction in the uncertainty
on the calculated value of the quark mass.

Nonperturbative renormalization of operators in lattice QCD using the
$\ARI$ (or $\RIp$)scheme has been successfully implemented for many
years. The evaluation of matrix elements in the $\RISMOM$
renormalization scheme in lattice simulations is equally practicable and
in view of the advantages explained above we strongly advocate its
use.\\

\vspace{2ex}
\noindent
{\bf Acknowledgments:}\\ We want to thank our colleagues in the
RBC-UKQCD Collaborations, especially Peter Boyle, for discussions and
encouragement.  C.T.S. warmly thanks Norman Christ and Bob Mawhinney for
their hospitality at Columbia University during the autumn term of
2008. This work was partially supported by U.S. DOE under Contract
No. DE-AC02-98CH10886 (A.S., C.S. and T.I.) and in part by RIKEN BNL
Research Center (T.I. and Y.A.). N.H.C was partially supported by the
U.S. DOE under Contract No. DE-FG02-92ER40699.  T.I. was also supported
in part by Grants-in-Aid for Scientific Research from the Ministry of
Education, Culture, Sports, Science and Technology No. 19740134 and No.
20025010, and by Japan Society for the Promotion of Science(JSPS) and
German Research Foundation(DFG), Japan-German Joint Research Project
2008-2009. C.T.S. acknowledges support from STFC Grant No. ST/G000557/1 and
EU Contract No. MRTN-CT-2006-035482 (Flavianet).
\begin{appendix}
\section{Alternative projectors for the vector and\\ axial-vector
operator Green's functions\label{app:AltProj}}
In general one can also use other projectors than those of the
$\RISMOM$ scheme as defined in Section~\ref{sec:GeneralNotations} in
order to define a scheme with a symmetric subtraction point. The general
structure before taking the trace with projectors of the one-loop
corrected amputated Green's functions of the operators in the
$\MSbar$ scheme for massless quarks with the momenta $p_1^2=p_2^2=q^2$
are given by
\begin{eqnarray}
\bar{\Lambda}_S&=&\left(
 \bA_{S}\*\unitop
+\bC_{S}\*{[\Fslash{p}_1,\Fslash{p}_2]\over q^2}
\right)\*\delta_{ij},\\
\bar{\Lambda}_P&=&\left(
 \bA_{P}\*i\*\gamma_5
+\bC_{P}\*i\*\gamma_5\*{[\Fslash{p}_1,\Fslash{p}_2]\over q^2}
               \right)\*\delta_{ij},\\
\bar{\Lambda}^{\mu}_V&=&\left(
 \bA_{V}\*\gamma^{\mu}
+\bB_{V}\*{\Fslash{p}_1\*\gamma^{\mu}\*\Fslash{p}_1
             +\Fslash{p}_2\*\gamma^{\mu}\*\Fslash{p}_2\over q^2}
+\bC_{V}\*{\Fslash{p}_1\*\gamma^{\mu}\*\Fslash{p}_2\over q^2}
+\bD_{V}\*{\Fslash{p}_2\*\gamma^{\mu}\*\Fslash{p}_1\over q^2}
\right)\*\delta_{ij},\\
\bar{\Lambda}^{\mu}_A&=&\left(
 \bA_{A}\*\gamma^{\mu}\*\gamma_5
-\bB_{A}\*{\Fslash{p}_1\*\gamma^{\mu}\*\gamma_5\*\Fslash{p}_1
             +\Fslash{p}_2\*\gamma^{\mu}\*\gamma_5\*\Fslash{p}_2\over q^2}
-\bC_{A}\*{\Fslash{p}_1\*\gamma^{\mu}\*\gamma_5\*\Fslash{p}_2\over q^2}
-\bD_{A}\*{\Fslash{p}_2\*\gamma^{\mu}\*\gamma_5\*\Fslash{p}_1\over q^2}
\right)\!\delta_{ij},\\
\bar{\Lambda}^{\mu\nu}_T&=&\left(
 \bA_T\*\sigma^{\mu\nu}
+\bB_T\*{\sigma^{\mu\nu}\*\Fslash{p}_2\*\Fslash{p}_1
       -\Fslash{p}_1\*\Fslash{p}_2\*\sigma^{\mu\nu}\over q^2}
+\bC_T\*{\Fslash{p}_1\*\Fslash{p}_2\*\sigma^{\mu\nu}\*
         \Fslash{p}_1\*\Fslash{p}_2\over q^4}
\right)\*\delta_{ij},
\end{eqnarray}
where the indices $i$ and $j$ denote color indices and the coefficient
functions read
\begin{eqnarray}
\bA_{S,P}&=& 1 + {\alpha_s\over4\*\pi}\*\CF\*\left[
             4
           + 3\*\log\left({\mu^2\over-q^2}\right)
           - {3\over2}\*C_0
 	  + \xi\*\left(
             2
           + \log\left({\mu^2\over-q^2}\right)
           - {C_0\over2}
 	  \right)
           \right],\\
\bC_{S,P}&=& {\alpha_s\over4\*\pi}\*\CF\*(1 - \xi)\*{C_0\over6},\\
\bA_{V,A}&=& 1 + {\alpha_s\over4\*\pi}\*\CF\*\left[ 
          - {C_0\over3}
          + \xi\*\left(
             1
           + \log\left({\mu^2\over-q^2}\right)
           - {C_0\over3}
           \right)
           \right],\\
\bB_{V,A}&=&-{\alpha_s\over4\*\pi}\*{\CF\over3}\*\left[
              (1 - \xi)\*(C_0 - 2) - 2
                            \right],\\
\bC_{V,A}&=&-{\alpha_s\over4\*\pi}\*{\CF\over3}\*\left[
            C_0 + 2 + \xi\*(C_0-1)
            \right],\\
\bD_{V,A}&=&{\alpha_s\over4\*\pi}\*{\CF\over3}\* \left[
            (1 - \xi)\*(C_0 - 1) - 1
           \right],\\
\bA_{T}&=& 1 - {\alpha_s\over4\*\pi}\*\CF\*(1-\xi)\*\left[
            {5\over 3}
          + \log\left({\mu^2\over -q^2}\right)
          - {2\over3}\*C_0
            \right],\\
\bB_{T}&=&-{\alpha_s\over4\*\pi}\*{\CF\over3}\* \left[
           2\*C_0 - (1-\xi)
         \right]
,\\
\bC_{T}&=& {\alpha_s\over4\*\pi}\*{\CF\over3}\*(1-\xi)\*\left[
            2 - C_0
         \right]
.
\end{eqnarray}
An example of a second possible choice for the projectors is the use of
the projectors of the $\ARI$ scheme for the amputated Green's function
of the vector and axial-vector operator
\begin{equation}
\label{eq:sRIVAR}
\lim_{m_R\to0}\left.{1\over48}\Tr\left[
\Lambda^{\mu}_{V,R}(p_1,p_2)\*\gamma_{\mu}
\right]\right|_{sym}\!\!\!\!\!=1,\qquad
\lim_{m_R\to0}\left.{1\over48}\Tr\left[
\Lambda^{\mu}_{A,R}(p_1,p_2)\*\gamma_5\*\gamma_{\mu}
\right]\right|_{sym}\!\!\!\!\!\!=1,
\end{equation}
in the renormalization conditions with a symmetric subtraction point
together with the conditions of Eqs.(\ref{eq:sRIPS}). One also 
has to modify the conditions of Eqs.(\ref{eq:RI})
\begin{eqnarray}
\label{eq:RIR}
\lim_{m_R\to0}\left.{1\over48}\left\{
\Tr\left[\gamma^{\mu}{\partial S_R^{-1}(p)\over\partial p^{\mu}}\right]\right|_{p^2=-\mu^2}
+\Tr\left.\left[
q_{\mu}\gamma^{\alpha}{\partial\over\partial q^{\alpha}}\Lambda^{\mu}_{V,R}
                                 \right]\right|_{sym}
\right\}
&=&-1,\\
\label{eq:sRISR}
\lim_{m_R\to0}{1\over12\*m_R}\*
\left\{
\left.\Tr\left[S^{-1}_R(p)\right]\right|_{p^2=-\mu^2}
-{1\over2}\*\left.\Tr\left[q_{\mu}\Lambda^{\mu}_{A,R}(p_1,p_2)\gamma_5\right]\right|_{sym}
\right\}
&=&\phantom{+}1
,
\end{eqnarray}
to maintain the Ward-Takahashi identities of
Eqs.(\ref{eq:vectorWI}) and (\ref{eq:axialWI}) for renormalized quantities.
This leads to a wave function renormalization factor $Z_q$ which is
different from the one of the $\ARI$ or $\RIp$ scheme. For this reason
the projectors used in Eqs.(\ref{eq:sRIVA}) of
Section~\ref{sec:GeneralNotations} have the advantage to produce the
same well-known renormalization constant $Z_q$ like in the
$\RIp$ scheme.  With the conditions of Eqs.(\ref{eq:sRIPS}) and
(\ref{eq:sRIVAR})-(\ref{eq:sRISR}) one obtains in this $\RISMOMB$ scheme
the following conversion factors:
\begin{equation}
C^{\rismomb}_q=1-{\alpha_s\over4\*\pi}\*C_F\*\left[
-1+{\xi\over2}\left(3-{2\over3}\Psi'\!\left({1\over3}\right)+\left({2\over3}\*\pi\right)^2\right)
\right]\!+\!\mathcal{O}\left(\alpha^2_s\right),
\end{equation}
\begin{equation}
\label{eq:conv}
C^{\rismomb}_P=1+{\als\over4\*\pi}\*\CF\*\!
\Bigg[
5+{2\over3}\*\pi^2-\Psi'\!\left({1\over3}\right)+{\xi\over2}
\Bigg]\!+\!\mathcal{O}\left(\alpha^2_s\right).
\end{equation}
The numerical evaluation of the resulting mass conversion factor in
the Landau gauge reads
\begin{equation}
\label{eq:num2}
C^{\rismomb}_m=1-{\als\over4\*\pi}\*C_F\*1.4841391\dots.
\end{equation}
\section{Anomalous dimensions\label{app:anomalous}}
In order to evaluate the mass in the $\RISMOM$ scheme at different scales
the mass anomalous dimension $\gamma^{\rismom}_m$ is required. It is
defined by
\begin{equation}
\label{eq:gammam}
\gamma_m={d\log m(\mu)\over d\log\left(\mu^2\right)}=
-\gamma^{(0)}_m{\alpha_s\over\pi}
-\gamma^{(1)}_m\*\left({\alpha_s\over\pi}\right)^2
+\mathcal{O}(\alpha^3_s).
\end{equation}
The result up to order $\alpha_s^2$ in the $\RISMOM$ scheme reads in
the Landau gauge
\begin{equation}
\label{eq:gammamrismom}
\gamma^{(0),\rismom}_m=\gamma^{(0),\msbar}_m,\qquad\qquad
\gamma^{(1),\rismom}_m=\gamma^{(1),\msbar}_m-{\beta^{(0)}\over4}\*\CF\*c^{(1),\rismom}_m(1,0),
\end{equation}
with $c^{(1),\rismom}_m(1,0)=-0.4841391...$ and the $\beta$ function
defined through
\begin{equation}
\label{eq:beta}
\beta={d\alpha_s(\mu)/\pi\over d\log(\mu^2)}=
-\beta^{(0)}\*\left({\alpha_s\over\pi}\right)^2
-\beta^{(1)}\*\left({\alpha_s\over\pi}\right)^3
+\mathcal{O}(\alpha^4_s).
\end{equation}
The first expansion coefficients for the mass anomalous dimension
in the $\MSbar$ scheme and the $\MSbar$ $\beta$ function are given by
\[
\begin{array}{rlrl}
\gamma_m^{(0),\msbar}&={3\over4}\*\CF,&
\gamma_m^{(1),\msbar}&={1\over16}\*\left({3\over2}\*\CF^2+{97\over6}\*\CF\*\CA-{10\over3}\*\CF\*\TF\*\nf\right),\\
\beta^{(0)}&={1\over4}\*\left({11\over3}\*\CA-{4\over3}\*\TF\*\nf\right),&&
\end{array}
\]
where $\CA$ is the Casimir operator in the adjoint representation of
SU(3) and $\nf$ is the number of active fermions. The symbol $\TF$
denotes the normalization of the trace of the SU(3) generators in 
the fundamental representation, conventionally chosen as $1/2$.\\
For the $\RISMOMB$ scheme, defined in Appendix~\ref{app:AltProj}, the
two-loop mass anomalous dimension is given by
\begin{equation}
\gamma^{(1),\rismomb}_m=\gamma^{(1),\msbar}_m-{\beta^{(0)}\over4}\*\CF\*c^{(1),\rismomb}_m,
\end{equation}
with $c^{(1),\rismomb}_m=-1.4841391\dots$ as given in Eq.~(\ref{eq:num2}).

Similarly the anomalous dimension $\gamma^{\rismom}_q=2{d\log\Psi\over
d\log(\mu^2)}$ can be defined, which is equal to the one in the
$\RIp$ scheme, $\gamma^{\rismom}_q=\gamma^{\rip}_q$, and can be found in
Refs.~\cite{Chetyrkin:1999pq,Gracey:2003yr} up to order $\alpha^3_s$. For
completeness we give here the result up to $\mathcal{O}(\alpha_s^2)$
which in the Landau gauge is the same as in the $\MSbar$ scheme
\begin{equation}
\gamma^{\rismom}_q= \left({\alpha_s\over\pi}\right)^2\*
\left({3\over32}\CF^2-{25\over64}\CF\*\CA+{1\over8}\CF\*\TF\*n_f\right)
	       +\mathcal{O}(\alpha_s^3).
\end{equation}
In the $\RISMOMB$ scheme, defined in Appendix~\ref{app:AltProj}, the
order $\alpha_s^2$ coefficient of the anomalous dimension
$\gamma^{\rismomb}_q$ in the Landau gauge reads
\begin{equation}
\gamma^{}_q= \left({\alpha_s\over\pi}\right)^2\*
\left({3\over32}\*\CF^2 - {31\over192}\*\CF\*\CA + {1\over24}\*\CF\*\TF\*n_f\right)
	       +\mathcal{O}(\alpha_s^3).
\end{equation}

We define the anomalous dimension $\gamma_T$ of the tensor
operator by 
\begin{equation}
\gamma^{\rismom}_T={d\log Z_{T}\over d\log(\mu^2)}=
 -\gamma_T^{(0)}\*\left({\alpha_s\over\pi}\right)
 -\gamma_T^{(1)}\*\left({\alpha_s\over\pi}\right)^2
 +\mathcal{O}(\alpha_s^3).
\end{equation}
In the $\RISMOM$ scheme it is given in the Landau gauge by
\begin{equation}
\gamma_{T}^{(0),\rismom}=\gamma_{T}^{(0),\msbar},\qquad\qquad
\gamma_{T}^{(1),\rismom}=\gamma_{T}^{(1),\msbar}-{\beta^{(0)}\over4}\*\CF\*c^{(1),\rismom}_{T}(0),
\end{equation}
where we have introduced the one-loop coefficient function
$c^{(1),\rismom}_{T}(\xi)$ extracted from Eq.(\ref{eq:CT}) using the
definition
$C^{\rismom}_T=1+{\als\over4\*\pi}\*\CF\*c^{(1),\rismom}_T(\xi)$, with
$c^{(1),\rismom}_{T}(0)=-0.1613797...$. The $\MSbar$ anomalous
dimension $\gamma_T^{\msbar}$ is known from
Refs.\cite{Broadhurst:1994se,Gracey:2000am,Gracey:2003yr} and reads
\begin{equation}
\gamma_{T}^{(0),\msbar}={1\over4}\*\CF,\qquad\qquad
\gamma_{T}^{(1),\msbar}={1\over16}\left(
- {19\over2}\*\CF^2
+ {257\over18}\*\CF\*\CA 
- {26\over9}\*\CF\*\TF\*\nf
                                 \right).
\end{equation}
Since the renormalization constants of the pseudoscalar and scalar
operator are related to the mass renormalization constant, the anomalous
dimensions of the pseudoscalar and scalar operator follow from the mass
anomalous dimension in Eq.(\ref{eq:gammamrismom}).
%
%
%
%
%
%
%
%
%

%
%
%
%
%
%
\end{appendix}

\end{document}